\documentstyle[12pt]{article}
\begin{document}
\thispagestyle{empty}
\newcommand{\p}[1]{(\ref{#1})}
\newcommand{\nn}{\nonumber}
\newcommand{\be}{\begin{equation}}
\newcommand{\ee}{\end{equation}}
\newcommand{\sect}[1]{\setcounter{equation}{0}\section{#1}}
\newcommand{\B}{{\cal B}}
\newcommand{\F}{{\cal F}}
\newcommand{\vs}[1]{\rule[- #1 mm]{0mm}{#1 mm}}
\newcommand{\hs}[1]{\hspace{#1mm}}
\newcommand{\mb}[1]{\hs{5}\mbox{#1}\hs{5}}
\newcommand{\Db}{{\overline D}}
\newcommand{\Fb}{{\overline F}}
\newcommand{\bea}{\begin{eqnarray}}
\newcommand{\eea}{\end{eqnarray}}
\newcommand{\wt}[1]{\widetilde{#1}}
\newcommand{\und}[1]{\underline{#1}}
\newcommand{\ov}[1]{\overline{#1}}
\newcommand{\sm}[2]{\frac{\mbox{\footnotesize #1}\vs{-2}}
           {\vs{-2}\mbox{\footnotesize #2}}}
\newcommand{\prt}{\partial}
\newcommand{\eps}{\epsilon}
\newcommand{\R}{\mbox{\rule{0.2mm}{2.8mm}\hspace{-1.5mm} R}}
\newcommand{\Z}{Z\hspace{-2mm}Z}
\newcommand{\cd}{{\cal D}}
\newcommand{\cg}{{\cal G}}
\newcommand{\ck}{{\cal K}}
\newcommand{\cw}{{\cal W}}
\newcommand{\vj}{\vec{J}}
\newcommand{\vl}{\vec{\lambda}}
\newcommand{\vz}{\vec{\sigma}}
\newcommand{\vt}{\vec{\tau}}
\newcommand{\vw}{\vec{W}}
\newcommand{\poiss}{\stackrel{\otimes}{,}}
\def\l#1#2{\raisebox{.2ex}{$\displaystyle
  \mathop{#1}^{{\scriptstyle #2}\rightarrow}$}}
\def\r#1#2{\raisebox{.2ex}{$\displaystyle
 \mathop{#1}^{\leftarrow {\scriptstyle #2}}$}}
\newcommand{\NP}[1]{Nucl.\ Phys.\ {\bf #1}}
\newcommand{\PL}[1]{Phys.\ Lett.\ {\bf #1}}
\newcommand{\NC}[1]{Nuovo Cimento {\bf #1}}
\newcommand{\CMP}[1]{Comm.\ Math.\ Phys.\ {\bf #1}}
\newcommand{\PR}[1]{Phys.\ Rev.\ {\bf #1}}
\newcommand{\PRL}[1]{Phys.\ Rev.\ Lett.\ {\bf #1}}
\newcommand{\MPL}[1]{Mod.\ Phys.\ Lett.\ {\bf #1}}
\newcommand{\BLMS}[1]{Bull.\ London Math.\ Soc.\ {\bf #1}}
\newcommand{\IJMP}[1]{Int.\ Jour.\ of\ Mod.\ Phys.\ {\bf #1}}
\newcommand{\JMP}[1]{Jour.\ of\ Math.\ Phys.\ {\bf #1}}
\newcommand{\LMP}[1]{Lett.\ in\ Math.\ Phys.\ {\bf #1}}
\renewcommand{\thefootnote}{\fnsymbol{footnote}}
\newpage
\setcounter{page}{0}
\pagestyle{empty}
\begin{flushright}
{November 1997}\\
{JINR E2-97-365}\\
{solv-int/9712002}
\end{flushright}
\vs{8}
\begin{center}
{\LARGE {\bf Extended $N=2$ supersymmetric}}\\[0.6cm]
{\LARGE {\bf matrix $(1,s)$-KdV hierarchies}}\\[1cm]

\vs{8}

{\large S. Krivonos$^{^1}$ and A. Sorin$^{2}$}
{}~\\
\quad \\
{\em {Bogoliubov Laboratory of Theoretical Physics, JINR,}}\\
{\em 141980 Dubna, Moscow Region, Russia}~\quad\\

\end{center}
\vs{8}

\centerline{ {\bf Abstract}}
\vs{4}

We propose the Lax operators for $N=2$ supersymmetric matrix
generalization of the bosonic $(1,s)$-KdV hierarchies. The
simplest examples -- the $N=2$ supersymmetric $a=4$
KdV and $a=5/2$ Boussinesq hierarchies -- are discussed in detail.

\vfill
{\em E-Mail:\\
1) krivonos@thsun1.jinr.dubna.su\\
2) sorin@thsun1.jinr.dubna.su }
\newpage

\pagestyle{plain}
\renewcommand{\thefootnote}{\arabic{footnote}}
\setcounter{footnote}{0}

\noindent{\bf 1. Introduction.}
The existence of three different infinite families of $N=2$ supersymmetric
integrable hierarchies
with the $N=2$ super $W_s$ algebras as their second Hamiltonian structure
is a well--established fact by now \cite{lm,y,yw}. Their bosonic limits
have been
analyzed in \cite{bks}, and three different families of the
corresponding bosonic hierarchies and their Lax
operators have been selected. Then, a complete description in terms of
super Lax operators for two out of three families has been proposed
in \cite{dg,bks1}, and the generalization to the matrix case
has been derived in \cite{bks1}.

The last remaining family of $N=2$ hierarchies is supersymmetrization
of the bosonic $(1,s)$-KdV hierarchies \cite{bks}. We call them the
$N=2$ supersymmetric $(1,s)$-KdV hierarchies. As opposed to the bosonic
counterparts of the former two hierarchies
\cite{bks}, the $(1,s)$-KdV hierarchy is irreducible (see \cite{bx} and
references  therein), i.e. its Lax operator cannot
be decomposed into a direct sum of some more elementary components.
This reduction property leads to a strong restriction of the
original supersymmetric Lax operator: its bosonic limit should
be irreducible. In other words, it should generate only a single operator
component. This property is surely satisfied for a supersymmetric Lax
operator which is a pure bosonic pseudo-differential operator
with the coefficients expressed in terms of $N=2$ superfields
and their fermionic derivatives in such a way that it commutes
with one of the two $N=2$ fermionic derivatives.
The Lax operator of this kind has in fact been observed in \cite{bks}
for the $N=2$ $a=5/2$ Boussinesq hierarchy in
the negative-power decomposition over bosonic
derivative up to the ${\partial}^{-5}$ order.
Quite recently, its closed analytic representation has been obtained
in \cite{gks}.

The aim of the present letter is to present a new infinite class of
reductions (with a finite number of fields) of $N=2$ supersymmetric
matrix KP hierarchy which includes the above--mentioned
family of $N=2$ $(1,s)$-KdV hierarchies in the scalar case.

{}~

\noindent{\bf 2. Extended matrix $N=2$ super $(1,s)$-KdV hierarchy.}
The Lax operator
\be
L_{KP}^{red}=I\partial+a_0+\omega_0 D+\sum_{j=-\infty}^{-1}
  \left( a_j\partial
 -\left[ Da_j\right] \Db+\omega_j D\partial-\frac{1}{2}
 \left[ D\omega_j\right]\left[ D,\Db\right]\right)\partial^{j-1} \; ,
\label{kpred}
\ee
derived by reduction $[D,L_{KP}^{red}]=0$ \cite{s} of the $N=2$
supersymmetric matrix KP hierarchy has been constructed in \cite{bks1}.
Here, $a_j$ and $\omega_j$ at $j \geq 1$  ($a_0$ and $\omega_0$) are
generic (chiral) bosonic and fermionic square matrix $N=2$ superfields.
The Lax operator \p{kpred} still contains
an infinite number of fields. Its further reductions \cite{bks1},
\be
(L^{red}_{KP})^s=I\partial^s+\sum_{j=1}^{s-1}\left( J_{s-j}\partial-
 \left[ DJ_{s-j}\right]\Db\right)\partial^{j-1}-J_s-
 \Db\partial^{-1}\left[ D J_s\right] - F\Fb -F\Db\partial^{-1}
 \left[ D\Fb\right] \; ,
\label{red1}
\ee
are characterized by a finite number of fields and contain two out of
three families of $N=2$ supersymmetric hierarchies
with $N=2$ super $W_s$ algebras as their second Hamiltonian structure
in the scalar case at $F={\overline F}=0$ (see \cite{bks1}
for details).

It appears that besides reductions \p{red1},
there exist other reductions of the Lax operator \p{kpred} which
in the scalar case correspond to the last remaining
family of $N=2$ hierarchies
with the $N=2$ super $W_s$ algebras as their second Hamiltonian structure,
i.e. $N=2$ $(1,s)$-KdV hierarchies.
Based on the inputs given above, we are led to the following
conjecture  for the expression of the matrix--valued pseudo--differential
operator with a finite number of superfields representing the new
reductions of the Lax operator \p{kpred}:
\begin{eqnarray}
L_{KP}^{red} \equiv
L_s= I\partial - \left[D {\cal L}_{s}^{-1}{\overline D}{\cal L}_{s}\right],
\quad {\cal L}_s \equiv I{\partial}^{s}+
{\sum}_{j=0}^{s-1}J_{s-j}{\partial}^{j}+
{\overline F}{\partial}^{-1}F.
\label{suplax}
\end{eqnarray}
Here, $s = 0, 1, 2, \ldots$,
$F\equiv F_{aA}(Z)$ and ${\overline F}\equiv {\overline F}_{Aa}(Z)$
($A,B=1,\ldots, k$; $a,b=1,\ldots , n+m$) are chiral and antichiral
rectangular matrix-valued $N=2$ superfields,
\begin{eqnarray}
D F=0, \quad {\overline D}~{\overline F} = 0,
\label{chiral}
\end{eqnarray}
respectively, and $J_j\equiv (J_j(Z))_{AB}$ are $k \times k$
matrix--valued bosonic $N=2$ superfields with the scaling dimensions
in length $[F]=[{\overline F}]=-(s+1)/2$,
$[J_j]=-j$; $I$ is the $k \times k$ unity matrix,
$I\equiv {\delta}_{A,B}$, and the matrix product is understood.
The matrix entries are bosonic superfields for $a=1,\ldots ,n$ and
fermionic superfields for $a=n+1,\ldots , n+m$, i.e.,
$F_{aA}{\overline F}_{Bb}=(-1)^{d_{a}{\overline d}_{b}}{\overline
F}_{Bb}F_{aA}$, where $d_{a}$ and ${\overline d}_{b}$ are the Grassmann
parities of the matrix elements $F_{aA}$ and ${\overline F}_{Bb}$,
respectively, $d_{a}=1$ $(d_{a}=0)$ for fermionic (bosonic) entries.
$Z=(z,\theta,\overline\theta)$ is a coordinate of
the $N=2$ superspace, $dZ \equiv dz d \theta d \overline\theta$.
In \p{suplax} the square brackets
mean that the $N=2$ supersymmetric fermionic covariant derivatives
$D$ and ${\overline D}$,
\begin{eqnarray}
D=\frac{\partial}{\partial\theta}
 -\frac{1}{2}\overline\theta\frac{\partial}{\partial z}, \quad
{\overline D}=\frac{\partial}{\partial\overline\theta}
 -\frac{1}{2}\theta\frac{\partial}{\partial z}, \quad
D^{2}={\overline D}^{2}=0, \quad
\left\{ D,{\overline D} \right\}= -\frac{\partial}{\partial z}
\equiv -{\partial},
\label{DD}
\end{eqnarray}
act only on the matrix superfields inside the brackets.
Let us stress the property of $L_s$ \p{suplax} to commute with the
fermionic derivative $D$, that is $[D,L_s]=0$.

The flows are the standard ones,
\begin{eqnarray}
{\textstyle{\partial\over\partial t_p}}L_s =
[A_p,L_s], \quad A_p=(L_s^{p})_{+},
\label{flows}
\end{eqnarray}
where $p = 1, 2, \ldots$, and
the subscript $+$ means a differential part of a pseudo-differential
operator. Let us remark that the Lax-pair representation \p{flows}
(with a generic matrix Lax operator of the type $[D,L]=0$) being
multiplied by the projector
$-D{\partial}^{-1}{\overline D}$ from the right,
can identically be rewritten in the same form
but with the operators $L$ and $A_p$ replaced by the operators
${\cal L} \equiv -D L{\partial}^{-1} {\overline D}$ and
${\cal A}_p \equiv -D A_p{\partial}^{-1} {\overline D}$, respectively,
obeying the chirality preserving condition
$D{\cal L}={\cal L}{\overline D}=0$ (as opposed to the former condition
$[D,L]=0$ we started with). In the scalar (generic matrix) case, Lax
operators of the last type have been considered in \cite{dg}
(\cite{bks1}). The relation of the chirality preserving scalar Lax
operators of Ref. \cite{dg} with the former ones (which have been
introduced in \cite{kst,bks,bks1}) was observed recently \cite{dg1}.

For the Lax operator $L_s$ \p{suplax}
the $N=2$ and $N=1$ residues\footnote{Let us recall that the standard $N=2$
($N=1$) super-residue is defined as the $N=2$ ($N=1$) superfield integral
of the coefficient of the operator $[D, {\overline D}]{\partial}^{-1}$
($D{\partial}^{-1}$ or ${\overline D}{\partial}^{-1}$).} vanish since it
does not contain the $N=2$ fermionic derivatives acting as operators.
Nevertheless, an infinite number of Hamiltonians can be obtained by using
the non--standard definition of the $N=2$ residue introduced in \cite{bks}
for the Lax operator of the $N=2$ $a=5/2$ Boussinesq hierarchy which
coincides with the definition of the residue for bosonic pseudo-differential
operators, i.e. it is the integrated coefficient of ${\partial}^{-1}$
\begin{eqnarray}
H_p= \int dz ~tr(L_s^{p})_{{\partial}^{-1}}|,
\label{res}
\end{eqnarray}
where $|$ means the $(\theta , {\overline {\theta}}) \rightarrow 0$ limit,
the integration is over the space coordinate $z$, and
$tr$ means the usual matrix trace.
These Hamiltonians are $N=2$ supersymmetric. Indeed, the operators
$tr(L_s^{p})_{{\partial}^{-1}}|$ (for the $L_s$ given by eqs. \p{suplax})
can be represented as
\begin{eqnarray}
tr(L_s^{p})_{{\partial}^{-1}}|=[D,{\overline D}]{\cal H}_p| +
\mbox{full space derivative terms}
\label{res2}
\end{eqnarray}
with local superfield functions ${\cal H}_p$. Consequently, the
Hamiltonians $H_p$ \p{res} can identically be rewritten in a manifestly
supersymmetric form
\begin{eqnarray}
H_p= \int dZ {\cal H}_p,
\label{res3}
\end{eqnarray}
where the integration is over the $N=2$ superspace coordinate $Z$.

One can easily derive the bosonic limit of the Lax operator $L_s$ \p{suplax}
at $F={\overline F}=0$,
\begin{eqnarray}
L_s^{bos}=
(I{\partial}^{s}+u_1{\partial}^{s-1}+
{\sum}_{j=0}^{s-2}u_{s-j}{\partial}^{j})^{-1}
(I{\partial}^{s+1}+u_1{\partial}^{s}+
{\sum}_{j=1}^{s-1}(u_{s-j+1}-v_{s-j}){\partial}^{j}-v_s),
\label{boslax}
\end{eqnarray}
where $u_j$ and $v_j$ are bosonic matrix components
of the superfield matrix $J_j$,
\begin{eqnarray}
u_j=J_j|, \quad v_j = D{\overline D} J_j|.
\end{eqnarray}
In the scalar case, i.e. at $k=1$,
the Lax operator $L_s^{bos}$ \p{boslax} in fact reproduces the Lax
operator $L^{(1)}_{[s;{\alpha}]}$ \cite{bks} of the $(1,s)$-KdV hierarchy.
Therefore, we conclude that the supersymmetric Lax operator $L_s$
\p{suplax} at $F={\overline F}=0$ corresponds to the matrix $N=2$
supersymmetric generalization of the bosonic $(1,s)$-KdV hierarchy, while
if it contains the superfield matrices $F$ and ${\overline F}$, it
generates the extended matrix $N=2$ $(1,s)$-KdV hierarchy.

{}~

\noindent{\bf 3. Examples: scalar case.}
To better understand what kind of hierarchies we have proposed,
let us discuss the first four hierarchies corresponding to the values
$s=0,1,2$ and $s=3$ in the Lax operator $L_s$ \p{suplax} in a simpler and
more studied scalar case (i.e., at $k=1$).
In this case, $J_j$ ($j=1,\ldots, s$) are generic scalar $N=2$ bosonic
superfields with spins $j$, while the chiral (antichiral) $N=2$
superfields $F_a$  ($\Fb_a$) contain $n$ bosonic and $m$
fermionic components with spin $(s+1)/2$.

{}~

\noindent{1. The $s=0$ case.}

For this simplest case the Lax operator \p{suplax}
\be \label{l0}
L_{0}=\partial -\left[ D \frac{1}{1+\Fb\partial^{-1}F}\Db
  (\Fb\partial^{-1}F)\right] \;
\ee
does not contain any
superfields $J_j$, and the chiral/antichiral superfields $F$ and $\Fb$ have
spins 1/2.
The second--flow equations \p{flows} have the following form:
\be \label{f0}
{\textstyle{\partial\over\partial t_2}}F=-F~''+2D(F\Fb~\Db F),\quad
{\textstyle{\partial\over\partial t_2}}\Fb=\Fb~''+
2\Db \left( (D\Fb )F\Fb \right)
\ee
and coincide with the corresponding flow of the $N=2$ GNLS hierarchy
\cite{bks}.
Therefore, the Lax operator \p{l0} provides a new description of the
$N=2$ GNLS hierarchy.

{}~

\noindent{2. The $s=1$ case.}

The Lax operator \p{suplax} has the following form:
\be
L_1= \partial - \left[ D \frac{1}{\partial +J_1 +\Fb\partial^{-1}F}
  \Db (J_1 +\Fb\partial^{-1}F) \right],
\label{scal1}
\ee
and the first two non-trivial flows \p{flows} read as
\begin{eqnarray}
&& {\textstyle{\partial\over\partial t_2}}{J_1} =
([D,{\overline D}~] J_1 - J_1^2+ 2{\overline F}F)~', \nonumber\\
&&{\textstyle{\partial\over\partial t_2}}F =
-F~''+ 2FD{\overline D}J_1, \quad
{\textstyle{\partial\over\partial t_2}}{\overline F} =
{\overline F}~''+ 2({\overline D} D J_1){\overline F},
\label{s1}
\end{eqnarray}
\bea
&& {\textstyle{\partial\over\partial t_3}}{J_1} =
 J_1~'''-3\left[ J_1[D,\Db ]J_1 -(\Db J_1)DJ_1-J_1^3+J_1\Fb F+
  (\Db F)D\Fb+\Fb F~'-\Fb~'F\right]', \nn \\
&&{\textstyle{\partial\over\partial t_3}}F = F~'''-3D\left[
\left( (\Db J)F\right)~'+J(\Db J)F-F\Fb~\Db F\right] ,\nn \\
&&{\textstyle{\partial\over\partial t_3}}{\overline F} =\Fb~'''+3\Db\left[
\left( (D J)\Fb\right)~'-J(DJ)\Fb+(D\Fb)F\Fb\right] \;.
\label{f1}
\eea
 From these expressions we can easily recognize that
after rescaling $J_1\rightarrow -2J_1,t_n\rightarrow -t_n$
they coincide with the corresponding flows  of the
$N=2$ $a=4$ KdV hierarchy \cite{lm}
at $F=\Fb=0$. With the non-zero superfields $F$ and
$\Fb$ we obtain a new extension of the $N=2$ $a=4$ KdV hierarchy.
Thus, our family of $N=2$ hierarchies
includes the well-known $N=2$ $a=4$ KdV hierarchy and possesses the
Lax-pair representation with the new Lax
operator\footnote{For alternative Lax-pair representations
of the $N=2$ $a=4$ KdV hierarchy, see Refs. \cite{lm,ks,kst,dg}.}
$L_1$ \p{scal1}.

{}~

\noindent{3. The $s=2$ case.}

This case is rather interesting because it corresponds to the
$N=2$ $a=5/2$ Boussinesq hierarchy which has been a puzzle
for a long time. The Lax operator \p{suplax}
\be
L_2= \partial - \left[ D \frac{1}{\partial^2 +J_1\partial+J_2 +
         \Fb\partial^{-1}F}
  \Db ( J_1\partial+J_2 +\Fb\partial^{-1}F) \right] \label{scal2}
\ee
gives rise to the following second--flow equations
\begin{eqnarray}
&& {\textstyle{\partial\over\partial t_2}}{J_1} =
(2J_2-J_1~'+2[D,{\overline D}~] J_1 - J_1^2)~',
\nonumber\\
&& {\textstyle{\partial\over\partial t_2}}{J_2} =
(J_2+2D{\overline D} J_1)~''+2({\overline F}F)'-2J_2 J_1~'+
2J_1D{\overline D} J_1~', \nonumber\\
&&{\textstyle{\partial\over\partial t_2}}F =
-F~''+ 2FD{\overline D}J_1, \quad
{\textstyle{\partial\over\partial t_2}}{\overline F} =
{\overline F}~''+ 2({\overline D} D J_1){\overline F}.
\label{f2}
\end{eqnarray}
In the new basis\footnote{The complexity
of these transformations is the price we have to pay for the simplicity
of the Lax operator \p{scal2}.},
\be\label{rescb}
t_2\rightarrow -t_2,\; J_1\rightarrow \frac{1}{3}J_1 ,\;
J_2\rightarrow -J_2+\frac{1}{2}J_1~'-\frac{1}{6} [D,\Db]J_1+\frac{2}{9}
{J_1}^2,
\ee
at $F=\Fb=0$, eqs. \p{f2} coincide with the $N=2$ $a=5/2$ Boussinesq
equation \cite{y}. Thus, we conclude that the $N=2$ $a=5/2$
Boussinesq hierarchy also belongs to the family of
$N=2$ super $(1,s)$-KdV
hierarchies with the Lax operator \p{suplax}.

{}~

\noindent{4. The $s=3$ case.}

As the last example, we present the second--flow equations
\begin{eqnarray}
&& {\textstyle{\partial\over\partial t_2}}{J_1} =
(2J_2-2J_1~'+3[D,{\overline D}~] J_1 - J_1^2)~', \nonumber\\
&& {\textstyle{\partial\over\partial t_2}}{J_2} =
(2J_3+J_2~'+6D{\overline D} J_1~')~'-2J_2J_1~'+
4J_1D{\overline D} J_1~', \nonumber\\
&& {\textstyle{\partial\over\partial t_2}}{J_3} =
(J_3+2D{\overline D} J_1~')~''+2J_1D{\overline D} J_1~''+
2J_2D{\overline D} J_1~'
\label{skdv4}
\end{eqnarray}
of the $N=2$ super $(1,3)$-KdV hierarchy which possesses the $N=2$ $W_4$
algebra as the second Hamiltonian structure.
This hierarchy contains the $N=2$ superfields $J_1$, $J_2$ and $J_3$ with
the spins $1$, $2$ and $3$, respectively, and its Lax operator looks like
\be
L_3= \partial - \left[D \frac{1}{\partial^3 +J_1\partial^2+J_2\partial +J_3}
  \Db( J_1\partial^2+J_2\partial + J_3) \right].
\label{scal3}
\ee
The extension of this system by the superfields $F$ and $\Fb$
can be straightforwardly derived from the Lax-pair representation
\p{suplax}, \p{flows}, and we do not present it here.

{}~

\noindent{\bf 4. Examples: matrix case.}
The construction of flows \p{flows} in the
matrix case goes without any new peculiarities.
The only difference with respect to the scalar case is
the appearance of some new terms in the flow equations and their ordering.
To demonstrate the difference, we present the Hamiltonian
densities\footnote{Let us recall that Hamiltonian densities are defined
up to terms which are fermionic or bosonic total derivatives of arbitrary
nonsingular, local functions of the superfield matrices.} and
the flow equations for the systems considered in the previous section
without comments.

{}~

\noindent{1. The $s=0$ case.}
\begin{eqnarray}
{\cal H}_2 = tr(\Fb F' + \Fb F\Fb F) \; ,
\label{h0}
\end{eqnarray}
\begin{eqnarray}
\label{mf0}
{\textstyle{\partial\over\partial t_2}}F=-F~''+2D(F\Fb~\Db F),\quad
{\textstyle{\partial\over\partial t_2}}\Fb=\Fb~''+
2\Db \left( (D\Fb )F\Fb \right).
\end{eqnarray}
These equations reproduce the second--flow equations of the $N=2$
supersymmetric matrix
GNLS hierarchy \cite{bks1}.

{}~

\noindent{2. The $s=1$ case.}
\begin{eqnarray}
{\cal H}_1= tr(J_1),\quad
{\cal H}_2= tr(\frac{1}{2}J_1^2-{\overline F}F),\quad
{\cal H}_3= tr(\frac{1}{3}J_1^3-J_1D{\overline D}J_1-{\overline F}F~'-
{\overline F}FJ_1),
\label{ham1}
\end{eqnarray}
\begin{eqnarray}
&& {\textstyle{\partial\over\partial t_2}}{J_1} =
([D,{\overline D}~] J_1 - J_1^2+ 2{\overline F}F)~'+
[J_1,[D,{\overline D}~] J_1], \nonumber\\
&&{\textstyle{\partial\over\partial t_2}}F =
-F~''+ 2FD{\overline D}J_1, \quad
{\textstyle{\partial\over\partial t_2}}{\overline F} =
{\overline F}~''+ 2({\overline D} D J_1){\overline F},
\label{kdv1}
\end{eqnarray}
\begin{eqnarray}
&&{\textstyle{\partial\over\partial t_3}}{J_1} =J_1~'''+3\left[\left(
 (\Db D J_1)J_1 - D(J_1\Db J_1)+\Fb~'F-\Fb F~'+D\Db (\Fb F)\right)~'-
 J_1D(J_1\Db J_1)\right. \nonumber \\
&& -\left. ( \Db (DJ_1)J_1)J_1+
 \left\{ J_1,D\Db (\Fb F )\right\}+\Fb F D\Db J_1+
   (\Db D J_1)\Fb F -J_1\Fb F~'-\Fb~' FJ_1 \right], \nonumber\\
&&{\textstyle{\partial\over\partial t_3}}F =
F~'''+ 3\left[D(F{\overline F}~{\overline D}F)- (FD{\overline D}J_1)~'-
FD(J_1{\overline D}J_1)\right], \nonumber\\
&&{\textstyle{\partial\over\partial t_3}}{\overline F} =
{\overline F}~'''+ 3{\overline D}\left[ (D{\overline F})F {\overline F}+
((DJ_1){\overline F})~'- (DJ_1)J_1{\overline F}\right],
\label{kdv2}
\end{eqnarray}
where the brackets ($\{,\}$) $[,]$ represent the (anti)commutator.

{}~

\noindent{3. The $s=2$ case.}
\begin{eqnarray}
{\cal H}_1= tr(J_1),\quad
{\cal H}_2= tr(\frac{1}{2}J_1^2-J_2),
\label{ham2}
\end{eqnarray}
\begin{eqnarray}
&& {\textstyle{\partial\over\partial t_2}}{J_1} =
(2J_2-J_1~'+2[D,{\overline D}~] J_1 - J_1^2)~'+[J_1,[D,{\overline D}~] J_1],
\nonumber\\
&& {\textstyle{\partial\over\partial t_2}}{J_2} =
(J_2+2D{\overline D} J_1)~''+2({\overline F}F)'-\{J_2,J_1~'\}+
2J_1D{\overline D} J_1~'+[J_2,[D,{\overline D}~]J_1], \nonumber\\
&&{\textstyle{\partial\over\partial t_2}}F =
-F~''+ 2FD{\overline D}J_1, \quad
{\textstyle{\partial\over\partial t_2}}{\overline F} =
{\overline F}~''+ 2({\overline D} D J_1){\overline F}.
\label{kdv3}
\end{eqnarray}

{}~

\noindent{4. The $s=3$ case.}
\begin{eqnarray}
{\cal H}_1= tr(J_1),\quad
{\cal H}_2= tr(\frac{1}{2}J_1^2-J_2) ,
\label{ham3}
\end{eqnarray}
\begin{eqnarray}
&& {\textstyle{\partial\over\partial t_2}}{J_1} =
(2J_2-2J_1~'+3[D,{\overline D}~] J_1 - J_1^2)~'+
[J_1,[D,{\overline D}~] J_1], \nonumber\\
&& {\textstyle{\partial\over\partial t_2}}{J_2} =
(2J_3+J_2~'+6D{\overline D} J_1~')~'-\{J_2,J_1~'\}+
4J_1D{\overline D} J_1~'+[J_2,[D,{\overline D}~]J_1], \nonumber\\
&& {\textstyle{\partial\over\partial t_2}}{J_3} =
(J_3+2D{\overline D} J_1~')~''+2J_1D{\overline D} J_1~''+
2J_2D{\overline D} J_1~'+ 2[J_3,D{\overline D}J_1].
\label{kdv4}
\end{eqnarray}

{}~

\noindent{\bf 5. Involution properties.}
Equations \p{mf0} and \p{kdv1}--\p{kdv2} admit the involution
\begin{eqnarray}
&& F^{*}= i^{s-1}{\cal I}{\overline F}^{T}, \quad
{\overline F^{*}}=i^{s-1}F^{T}{\cal I}, \quad
J_j^{*}=(-1)^{j}J_j^{T}, \nonumber\\
&& {\theta}^{*}={\overline {\theta}}, \quad
{\overline {\theta}^{*}}={\theta}, \quad
t^{*}_p= (-1)^{p+1} t_p, \quad z^{*}= z, \quad i^{*}=-i,
\label{conjs}
\end{eqnarray}
for $s=0$ and $s=1$, respectively. Here, $i$ is the imaginary unit, the
symbol $T$ means the operation of matrix transposition, and the matrix
${\cal I}$ is
\begin{eqnarray}
{\cal I}\equiv (-i)^{d_a} {\delta}_{ab}, \quad
{\cal I}{\cal I}^{*}=I, \quad {\cal I}^{3}={\cal I}^{*},\quad
{\cal I}^2= (-1)^{d_a}{\delta}_{ab}.
\label{matrII}
\end{eqnarray}
The same involution property is not
straightforwardly satisfied for eqs. \p{kdv3} ($s=2$) and eqs. \p{kdv4}
($s=3$): it is satisfied
in a new basis with the superfields $J_2$ and $J_2, J_3$
being replaced by
\begin{eqnarray}
J_2 \Rightarrow J_2 -\frac{1}{2}J_1~',
\label{trn}
\end{eqnarray}
and
\begin{eqnarray}
J_2 \Rightarrow J_2 -J_1~',\quad
J_3 \Rightarrow J_3 -\frac{1}{2} J_2~',
\label{trn1}
\end{eqnarray}
respectively, while all the other superfields are unchanged.
It seems reasonable to conjecture the existence of a basis in the space of
superfield matrices where the involution \p{conjs} is admitted for
any given value of the parameter $s$ that parametrizes the Lax
operator $L_s$ \p{suplax}.

{}~

\noindent{\bf 6. Conclusion. }
In this letter, we constructed a new infinite variety of matrix $N=2$
supersymmetric hierarchies by exhibiting the corresponding super Lax
operators. Their involution properties are analyzed.
As a byproduct, we solved the problem of a Lax-pair description for the
last remaining family of $N=2$ hierarchies with the $N=2$ super $W_s$
algebras as their second Hamiltonian structure and
derived new extensions of such
familiar hierarchies as the $N=2$ supersymmetric $a=4$ KdV and $a=5/2$
Boussinesq hierarchies. New bosonic hierarchies can be obtained from the
constructed supersymmetric counterparts in the bosonic limit.

{}~

\noindent{\bf Acknowledgments.}
We would like to thank L. Bonora for his interest in this paper
and hospitality at SISSA. This work was partially supported by the Russian
Foundation for Basic Research, Grant No. 96-02-17634, RFBR-DFG Grant No.
96-02-00180, INTAS Grant No. 93-127 ext..

\end{document}